\documentstyle[amssymb,prl,aps,epsfig]{revtex}

\begin{document}
\draft
\twocolumn[\hsize\textwidth\columnwidth\hsize\csname@twocolumnfalse\endcsname

\title{Thermodynamic relevance of nanoscale inhomogeneities in Bi$_{2}$Sr$_{2}$CaCu$_{2}$O$_{8+\delta }$}
\author{T. Schneider \\
Physik-Institut der Universit\"{a}t Z\"{u}rich, Winterthurerstrasse 190,\\
CH-8057 Z\"{u}rich, Switzerland}
\maketitle
\begin{abstract}
Using finite size scaling we analyze specific heat and London
penetration depth data to estimate the spatial extent of the
superconducting grains in so called high-quality
Bi$_{2}$Sr$_{2}$CaCu$_{2}$O$_{8+\delta }$ single crystals.
Contrary to the previously investigated type II superconductors,
YBa$_{2}$Cu$_{3}$O$_{7-\delta }$, MgB$_{2}$, 2H-NbSe$_{2}$ and
Nb$_{77}$Zr$_{23}$, where the lower bound for the length scale of
the grains ranges from $182$ A to $814$ A, our analysis uncovers
nanoscale superconducting grains. This clarifies the relevance of
the spatial variations in the electronic characteristics observed
in underdoped Bi$_{2}$Sr$_{2}$CaCu$_{2}$O$_{8+\delta }$ with
scanning tunnelling microscopy for bulk and thermodynamic
properties and establishes their spatial extent.
\bigskip
\bigskip
\end{abstract}
]

Since the discovery of superconductivity in cuprates by Bednorz
and M\"{u}ller\cite{bed} a tremendous amount of work has been
devoted to their characterization. Bismuth based cuprates deserved
special attention due to their potential for applications. The
issue of inhomogeneities and their characterization is essential
for several reasons, including: First, if inhomogeneity is an
intrinsic property, a re-interpretation of experiments, measuring
an average of the electronic properties, is unavoidable. Second,
inhomogeneity may point to a microscopic phase separation, i.e.
superconducting grains, embedded in a nonsuperconducting matrix.
Third, there is neutron spectroscopic evidence for nanoscale
cluster formation and percolative superconductivity in various
cuprates\cite{mesot,furrer}. Fourth, nanoscale spatial variations
in the electronic characteristics have been observed in underdoped
Bi$_{2}$Sr$_{2}$CaCu$_{2}$O$_{8+\delta }$ with scanning tunnelling
microscopy (STM)\cite{liu,chang,cren,lang}. They reveal a spatial
segregation of the electronic structure into 3nm diameter
superconducting domains in an electronically distinct background.
On the contrary, a large degree of homogeneity has been observed
by Renner and Fischer\cite{renner}. As STM is a surface probe the
relevance of these observations for bulk and thermodynamic
properties remains to be clarified. Fifth, in
YBa$_{2}$Cu$_{3}$O$_{7-\delta }$, MgB$_{2}$, 2H-NbSe$_{2}$ and
Nb$_{77}$Zr$_{23}$ considerably larger grains have been
established. The magnetic field induced finite size effect
revealed lower bounds ranging from $L=182$ to $814$A \cite{tsfs}.
This letter concentrates on the Bismuth- and Thallium based
cuprates and addresses these issues by providing clear evidence
for the existence of superconducting grains with spatial nanoscale
extent.

It is well-known that systems of finite extent, i.e. isolated
superconducting grains, undergo a rounded and smooth phase
transition\cite {fisher}. As in an infinite and homogeneous system
the transition temperature $T_{c}$ is approached the correlation
length $\xi $ increases strongly and diverges at $T_{c}$. However,
when superconductivity is restricted to grains with length scale
$L$, $\xi $ cannot grow beyond $L$. In type II superconductors,
exposed to a magnetic field $H$, there is an additional limiting
length scale $L_{H}=\sqrt{\Phi _{0}/\left( aH\right) }$ with
$a\approx 3.12$, related to the average distance between vortex
lines \cite{tsfs}. Indeed, as the magnetic field increases, the
density of vortex lines becomes greater, but this cannot continue
indefinitely, the limit is roughly set on the proximity of vortex
lines by the overlapping of their cores. Due to these limiting
length scales the phase transition is rounded and occurs smoothly.
As a remnant of the singularity in the thermodynamic quantities of
the homogeneous system at $T_{c}$, these quantities exhibit a so
called finite size effect, i.e. a maximum or an inflection point
at $T_{p} $, where $\xi \left( T=T_{p}\right) =L_{m}$ and $L_{m}$
$=L$ when $L<L_{H}$, while $L>L_{H}$ for $L_{m}=L_{H}$. We have
shown that in YBa$_{2}$Cu$_{3}$O$_{7-\delta }$, MgB$_{2}$,
2H-NbSe$_{2}$ and Nb$_{77}$Zr$_{23}$, in the range of attained
fields, the condition $L>L_{H}$ is satisfied. As the experimental
data do not extend to low fields, the actual extent of the grains
may exceed the resulting lower bounds ranging from $L=182$ to
$814$A \cite{tsfs}. On the contrary, the specific heat
measurements reveal that in the Bismuth- and Thallium based
cuprates $T_{p}$ does not shift up to $14$T, for fields applied
parallel or perpendicular to the c-axis \cite
{junod,mirmelstein,junodbi2212}. Thus, in these materials the
condition $L<L_{H}=$ $\sqrt{\Phi _{0}/\left( aH\right) }=69$A
applies, uncovering nanoscale superconducting grains, consistent
with the length scale of the inhomogeneities observed with STM
spectroscopy\cite{liu,chang,cren,lang}.

To refine this rough estimate we perform a detailed finite size
scaling analysis of the specific heat \cite{junodbi2212} and the
London penetration depth data \cite{jacobs} of
Bi$_{2}$Sr$_{2}$CaCu$_{2}$O$_{8+\delta }$ single crystals. It
uncovers isolated superconducting grains with length scale
$L_{c}\approx 92$A along the c-axis and $L_{ab}\approx 68$A in the
ab-plane. The latter value is comparable with the spatial
segregation of the electronic structure into 30A diameter
superconducting domains in an electronically distinct background
observed with STM spectroscopy\cite {liu,chang,cren,lang}. As STM
is a surface probe, our analysis establishes that these grains are
not merely an artefact of the surface, but a bulk property with
spatial extent, giving rise to a rounded thermodynamic phase
transition which occurs smoothly. While there are many questions
to be answered about the intrinsic or extrinsic origin of the
inhomogeneity, the existence and the nature of a macroscopic
superconducting state, we establish that a finite size scaling
analysis yields spatial extent of the inhomogeneity which prevents
the occurrence of a homogeneous and macroscopic superconducting
phase.

Approaching $T_{c}$ from below the transverse correlation length
$\xi _{i}^{t}$ in direction $i$ increases strongly. However, due
to the limiting length scales $L_{i}$ and $L_{H_{i}}=\sqrt{\Phi
_{0}/\left( aH_{i}\right) }$, it is bounded and cannot grow beyond
\begin{equation}
\xi _{i}^{t}\left( t_{p}\right) =\xi _{0i}^{t}\left| t_{p}\right|
^{-\nu }=\left\{
\begin{array}{c}
\sqrt{L_{j}L_{k}},\ i\neq j\neq k \\
\sqrt{\Phi _{0}/\left( aH_{i}\right) }=L_{H_{i}}
\end{array}
\right\} ,  \label{eq1}
\end{equation}
where $t_{p}=1-T_{p}/T_{c}$ and $\nu $ denotes the critical
exponent of the transverse correlation length. Beyond the
mean-field approximation it differs from $\nu =1/2$ and $a\approx
3.12$ is a universal constant\cite {tsfs}. As a remnant of the
singularity at $T_{c}$ the thermodynamic quantities exhibit a so
called finite size effect, i.e. a maximum or an inflection point
at $T_{p}$. Two limiting regimes, characterized by
\begin{equation}
L_{H_{i}}=\sqrt{\frac{\Phi _{0}}{aH_{i}}}=\left\{
\begin{array}{c}
<\sqrt{L_{j}L_{k}} \\
>\sqrt{L_{j}L_{k}}
\end{array}
,i\neq j\neq k\right\} ,  \label{eq2}
\end{equation}
can be distinguished. For $L_{H_{i}}<\sqrt{L_{j}L_{k}}$ the
magnetic field induced finite size effect limits $\xi _{i}^{t}$ to
grow beyond $L_{H_{i}}$, while for $L_{H_{i}}>\sqrt{L_{j}L_{k}}$
the superconducting grains set the limiting length scales. Since
$L_{H_{i}}$ can be tuned by the strength of the applied magnetic
field, both limits are experimentally accessible.
$L_{H_{i}}<\sqrt{L_{j}L_{k}}$ is satisfied for sufficiently high
and $L_{H_{i}}>\sqrt{L_{j}L_{k}}$ for low magnetic fields. Thus,
the occurrence of a magnetic field induced finite size effect
requires that the magnetic field and the length scales of the
superconducting grains satisfy the lower bound $H_{i}>\Phi
_{0}/\left( aL_{i}L_{j}\right) $.

\begin{figure}
\centering
\includegraphics[width= 0.95\linewidth]{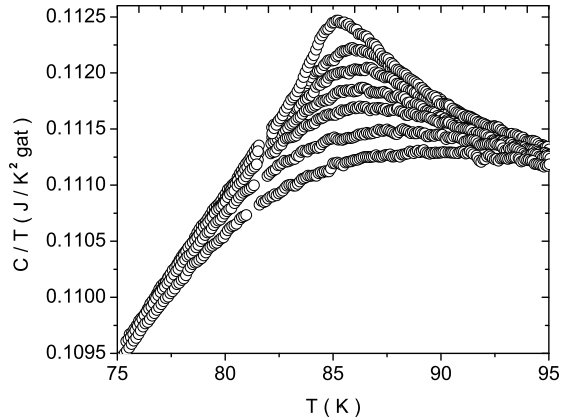} \vskip12pt
\caption{Specific heat coefficient of a
Bi$_{2.12}$Sr$_{1.71}$Ca$_{1.22}$Cu$_{1.95}$O$_{8+\delta }$ single
crystal versus temperature taken from Junod \emph{et
al.}\protect\cite{junodbi2212}. The magnetic field is parallel to
the c-axis (from top to bottom: $0,0.5,1,2,4,8$ and $14$T.}
\label{fig1}
\end{figure}

To derive detailed estimates we analyze the single crystal data
for the specific heat coefficient of Junod \emph{et
al}.\cite{junodbi2212} for
Bi$_{2.12}$Sr$_{1.71}$Ca$_{1.22}$Cu$_{1.95}$O$_{8+\delta }$. In
Fig.\ref{fig1} we displayed the data for fields applied parallel
to the c-axis. In zero field there is a broad peak around $T_{p}$
$=85.1$K, establishing the existence of superconducting grains.
Furthermore, $T_{p}$ does not appear to shift with increasing
magnetic field up to $H=14$T. Indeed, a background independent
definition of $T_{p}$ is the temperature at which $\left( c\left(
H,T\right) -c\left( H=0,t\right) \right) /T$ is minimum. This
definition leads to the conclusion that $T_{p}$ does not change up
to $14$T \cite{junodbi2212}. Noting that the same behavior was
found for fields perpendicular to the c-axis, inequality
(\ref{eq2}) yields for the spatial extent of the superconducting
grains the upper bounds
\begin{equation}
L_{ab}<69A,\ \sqrt{L_{ab}L_{c}}<69A.  \label{eq3}
\end{equation}
Another striking feature is the reduction of the peak height with
increasing field, applied parallel to the c-axis, while for fields
applied parallel to the layers there is a minor reduction
only\cite{junodbi2212}.

\begin{figure}
\centering
\includegraphics[width= 0.95\linewidth]{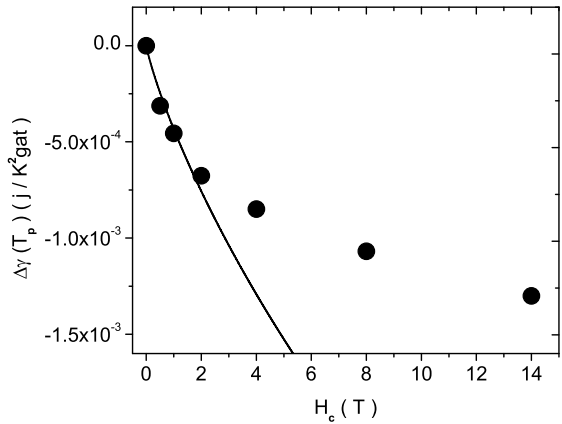} \vskip12pt
\caption{$\Delta \gamma \left( t_{p},H_{c}\right) $ versus
magnetic field $H_{c}$ applied parallel to the c-axis of a
Bi$_{2.12}$Sr$_{1.71}$Ca$_{1.22}$Cu$_{1.95}$O$_{8+\delta }$ single
crystal, derived from Junod \emph{et al.}
\protect\cite{junodbi2212}. The solid line is Eq.(\ref{eq5}) with
the parameters listed in the text.} \label{fig2}
\end{figure}

To elucidate this behavior we invoke close to $T_{c}$ the scaling
form\cite{tsfs,book}
\begin{equation}
\frac{c\left( t,H_{i}\right) }{T}=\frac{c\left( t\right)
}{T}G\left( z\right) +B^{\pm },\ \ z=\frac{H_{i}\left( \xi
_{i}^{t}\right) ^{2}}{\Phi _{0}},  \label{eq4}
\end{equation}
where
\begin{equation}
\ \frac{c\left( t\right) }{T}=\frac{A^{\pm }}{_{\alpha }}\left|
t\right| ^{-\alpha }  \label{eq4a}
\end{equation}
is the zero field contribution. $G\left( z\right) $ is a universal
scaling function of its argument and $G\left( 0\right)
=1$\cite{tsfs,book}. $\alpha $ denotes the critical exponent and
$A^{\pm }$ the critical amplitude of the specific heat, where $\pm
=sign(1-T_{c}/T)$ and $B^{\pm }$ accounts for the regular
background. Strictly speaking this scaling form applies to a
neutral superfluid. However, there is strong evidence that for the
accessible temperature range cuprates fall effectively into this
3D-XY universality class \cite{tsfs,book,osborn}. At $T_{p}$ the
transverse correlation length is fixed by $\xi _{i}^{t}\left(
t_{p}\right) =\sqrt{L_{j}L_{k}}$(Eq.(\ref{eq1})). Accordingly,
$H_{i}\rightarrow 0$ corresponds to limit $z\rightarrow 0$. Here
the scaling function $G\left( z\right) $ adopts the
form\cite{book}, $G\left( z\right) =1+cz\left( -1+\ln \left(
z\right) \right) $, where $c$ is a universal constant. At $T_{p}$
and for fields applied along the c-axis the relative reduction of
the specific heat coefficient, $\Delta \gamma \left(
t_{p},H_{c}\right) =\left( c\left( t_{p},H_{c}\right) -c\left(
t_{p},H_{c}=0\right) \right) /T_{p}$, adopts then the scaling form
\begin{equation}
\Delta \gamma \left( t_{p},H_{c}\right) =bH_{c}\left( -1+\ln
\left( dH_{c}\right) \right) ,  \label{eq5}
\end{equation}
where
\begin{equation}
b=\frac{A^{-}cL_{a}L_{b}}{T_{p}\alpha \Phi _{0}}\left|
t_{p}\right| ^{-\alpha },\ \ dH_{c}=z=\frac{H_{c}L_{a}L_{b}}{\Phi
_{0}}.  \label{eq5a}
\end{equation}
Hence the reduction is controlled by the length scales of the
superconducting grains in the ab-plane. As this scaling form
applies in the limit $H_{c}\rightarrow 0$, we restricted the fit
to the experimental data displayed in Fig.\ref{fig2} to the low
field data points. The fit yields $b=8\ 10^{-5}$J/$\left(
\text{K}^{2}\text{gatT}\right) $and$\ d=0.012$ T$^{-1} $ so that
\begin{equation}
\sqrt{L_{a}L_{b}}=50\text{A,}  \label{eq6}
\end{equation}
consistent with the upper bound, $L_{ab}<69$A (Eq.(\ref{eq3})),
derived from the absence of a shift of $T_{p}$ up to $14$T. When
the magnetic field is applied parallel to the a-axis, the scaling
variable is $z=\left( H_{a}\left( \xi _{a}^{t}\right) ^{2}\right)
/\Phi _{0}\approx \left( H_{a}\left( \xi _{ab}^{t}\right)
^{2}\right) /\Phi _{0}=\left( H_{a}\left( \xi _{c}^{t}/\gamma
^{2}\right) ^{2}\right) /\Phi _{0}$, where $\gamma $, defined by
$\gamma ^{2}=\xi _{c}^{t}/\xi _{ab}^{t}=\left( \lambda
_{c}/\lambda _{ab}\right) ^{2}$, measures the anisotropy. Thus, in
this field direction the scaling variable entering Eq.(\ref{eq5})
is $z=H_{ab}L_{ab}^{2}/\left( \gamma ^{4}\Phi _{0}\right) $. As
Bi$_{2}$Sr$_{2}$CaCu$_{2}$O$_{8+\delta }$ is highly anisotropic
($\gamma >100$) the reduction should be very small. This behavior,
reflecting the large anisotropy, was also found
experimentally\cite{junodbi2212}.

\begin{figure}
\centering
\includegraphics[width= 0.95\linewidth]{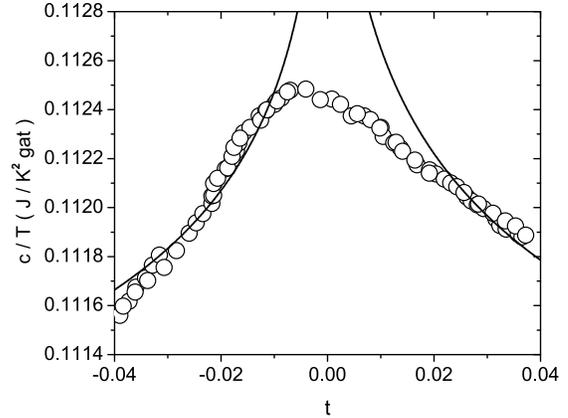} \vskip12pt
\caption{Zero field specific heat coefficient of a
Bi$_{2.12}$Sr$_{1.71}$Ca$_{1.22}$Cu$_{1.95}$O$_{8+\delta }$ single
crystal derived from Junod \emph{et
al.}\protect\cite{junodbi2212}. The solid lines are
Eq.(\ref{eq4a}) with the parameters listed in the text.}
\label{fig3}
\end{figure}

Next we turn to the zero field transition to estimate the critical
correlation volume $V_{c}^{t}=\xi _{0a}^{t}\xi _{0b}^{t}\xi
_{0c}^{t}$. This quantity is essential to derive, in combination
with the volume of the superconducting grains,
$V_{gr}=L_{a}L_{b}L_{c}$, an independent estimate of the reduced
temperature $t_{p}$, where the finite size effect sets in. In
Fig.\ref{fig3} we compare the measured temperature dependence of
the zero field specific heat coefficient with the expected
critical behavior of the homogenous counterpart. The solid lines
in Fig.\ref{fig3} are Eqs.(\ref{eq4}) and (\ref{eq4a}) with $H=0$
and the parameters $T_{c}=85.7$K, $\widetilde{A}^{-}=A^{-}/\left(
T_{c}\alpha \right) =-0.0472$,$\ B^{-}=0.15693$, $B^{+}=0.16022$
in J/K$^{2}$gat, $A^{+}/A^{-}=1.07$ and $\alpha =-0.013$.

To estimate the correlation volume we note that $A^{-}=T_{c}\alpha
$ $\widetilde{A}^{-}$, $A^{-}$(cm$^{-3}$)$=\left(
10^{4}/k_{B}/V_{gat}\right) A^{-}$(mJ/K/cm$^{\text{3}}$) and
$V_{gat}=9.1$cm$^{3}$ gives for the critical amplitude of the
specific heat the estimate $A^{-}=4.2\ 10^{-3}$A$^{-3}$. The
universal relation\cite{tsfs}, $A^{-}\xi _{0a}^{t}\xi _{0b}^{t}\xi
_{0c}^{t}=A^{-}V_{c}^{t}=\left( R^{-}\right) ^{3}$, where$\
R^{-}=0.815$, yields then the correlation volume $V_{c}^{t}\approx
129$A$^{3} $ compared to $680$A$^{3}$, the value for optimally
doped YBa$_{2}$Cu$_{3}$O$_{7-\delta }$\cite{tsfs}.

\begin{figure}
\centering
\includegraphics[width= 0.95\linewidth]{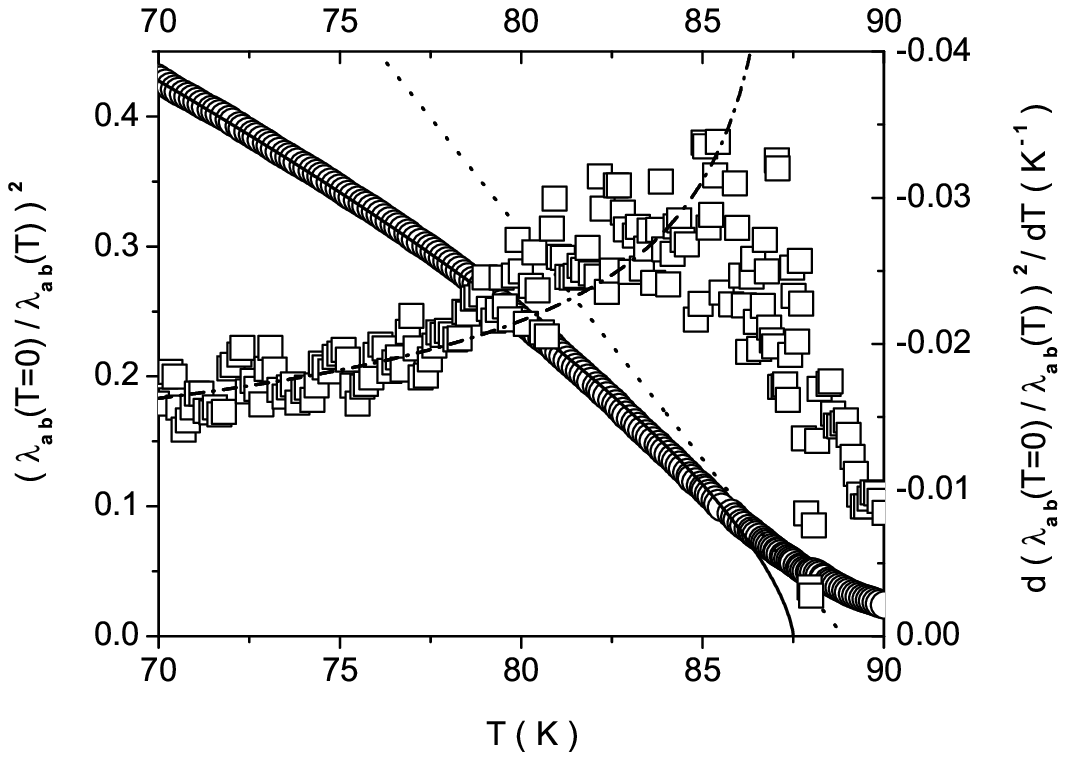} \vskip12pt
\caption{Microwave surface impedance data for $\lambda
_{ab}^{2}\left( T=0\right) /\lambda _{ab}^{2}\left( T\right) $
($\bigcirc $)\ and $d\left( \lambda _{ab}^{2}\left( T=0\right)
/\lambda _{ab}^{2}\left( T\right) \right) /dT$ ($\square $) versus
$T$ of a high-quality Bi$_{2}$Sr$_{2}$CaCu$_{2}$O$_{8+\delta }$
single crystal taken from Jacobs \emph{et al.}
\protect\cite{jacobs}. The solid line is $\lambda _{ab}^{2}\left(
T=0\right) /\lambda _{ab}^{2}\left( T\right) =1.2\left(
1-T/T_{c}\right) ^{2/3}$ and the dash-dot line its derivative with
$T_{c}=87.5$K, indicating the leading critical behavior of the
homogeneous system. The dotted line is the tangent to the
inflection point at $T_{p}\approx 87$K, where $d\left( \lambda
_{ab}^{2}\left( T=0\right) /\lambda _{ab}^{2}\left( T\right)
\right) /dT$ is maximum.} \label{fig4}
\end{figure}

Finally we turn to the finite size effect in the London
penetration depth. Considering again the 3D-XY critical point,
extended to the anisotropic case, the penetration depths and
transverse correlation lengths in directions $i$ and $j$ are
universally related by\cite{book,hohenberg}
\begin{equation}
\frac{1}{\lambda _{i}\left( T\right) \lambda _{j}\left( T\right)
}=\frac{16\pi ^{3}k_{B}T}{\Phi _{0}^{2}\sqrt{\xi _{i}^{t}\left(
T\right) \xi _{j}^{t}\left( T\right) }}.  \label{eq9}
\end{equation}
When the superconductor is inhomogeneous, consisting of
superconducting grains with length scales $L_{i}$, embedded in a
nonsuperconducting medium,\ $\xi _{i}^{t}$ does not diverge but is
bounded by $\xi _{i}^{t}\xi _{j}^{t}\leq L_{k}^{2}$, where $i\neq
j\neq k$. The resulting finite size effect is clearly seen in the
microwave surface impedance data for $\lambda _{ab}^{2}\left(
T=0\right) /\lambda _{ab}^{2}\left( T\right) $ versus $T$ of
Jacobs \emph{et al.}\cite{jacobs},\ displayed in Fig.\ref{fig4}.
The solid curve indicates the leading critical behavior of the
homogeneous system. A characteristic feature of this finite size
effect is the occurrence of an inflection point at $T_{p}\approx
87$K, giving rise to an extremum in $d\left( \lambda
_{ab}^{2}\left( T=0\right) /\lambda _{ab}^{2}\left( T\right)
\right) /dT$. Here Eq.(\ref{eq9}) reduces to
\begin{equation}
\frac{1}{\lambda _{ab}^{2}\left( T_{p}\right) }\approx
\frac{1}{\lambda _{a}\left( T_{p}\right) \lambda _{b}\left(
T_{p}\right) }=\frac{16\pi ^{3}k_{B}T_{p}}{\Phi _{0}^{2}L_{c}}.
\label{eq10}
\end{equation}
With $\lambda _{ab}\left( T=0\right) =1800$A as obtained from $\mu
$SR measurements \cite{leem}, $T_{p}\approx 87$K and $\lambda
_{ab}^{2}\left( T=0\right) /\lambda _{ab}^{2}\left( T_{P}\right)
=0.066$ we find
\begin{equation}
L_{c}\approx 68\text{A,}  \label{eq11}
\end{equation}
compared to $L_{ab}\approx 50$A (Eq.(\ref{eq6})) and consistent
with the upper bound $\sqrt{L_{ab}L_{c}}<69$A (Eq.(\ref{eq3})).
This yields for the
volume of the superconducting grains, $V_{gr}$, the estimate $%
V_{gr}=L_{ab}^{2}L_{c}\approx 1.7\ 10^{5}$A$^{3}$. Together with
the correlation volume $V_{c}^{t}\approx 129$A$^{3}$ we find for
the reduced temperature $t_{p}$, where rounding and shifting of
the transition sets in, the value $\left| t_{p}\right| =\left(
V_{c}^{t}/V_{gr}\right) ^{1/3\nu }\approx \left(
V_{c}^{t}/V_{gr}\right) ^{1/2}$ $\approx 0.027$, consistent with
the rounding of the peak in the specific heat coefficient
displayed in Fig.\ref{fig3}.\ Thus, although the superconducting
grains are of nanoscale only, the very small correlation volume
makes the critical regime attainable. Clear evidence for 3D-XY
critical behavior was also observed in epitaxially grown Bi2212
films\cite{osborn}.

In contrast to YBa$_{2}$Cu$_{3}$O$_{7-\delta }$, MgB$_{2}$,
2H-NbSe$_{2}$ and Nb$_{77}$Zr$_{23}$, where the lower bounds for
the length scale $L$ of the superconducting grains ranges from
$182$ A to $814$ A \cite{tsfs}, we have seen that the data for
Bi$_{2}$Sr$_{2}$CaCu$_{2}$O$_{8+\delta }$ single crystals uncovers
the existence of nanoscale superconducting grains. This finding is
consistent with the spatial variations in the electronic
characteristics observed in underdoped Bi-2212 with STM\cite
{liu,chang,cren,lang}. However, our analysis has shown that
nanoscale superconducting grains are not merely an artefact of the
surface, but a bulk property with spatial extent, giving rise to
finite size effects and with that to a rounded thermodynamic phase
transition which occurs smoothly. While there are many questions
to be answered about the intrinsic or extrinsic origin of the
inhomogeneity, the existence and the nature of a macroscopic
superconducting state, we established that a finite size scaling
analysis yields the spatial extent of the inhomogeneity which
prevents the occurrence of a homogeneous and macroscopic
superconducting phase.

The author is grateful to A. Furrer, A. Junod, H. Keller, M.
Kugler, J. Loram and K.A. M\"{u}ller for very useful comments and
suggestions on the subject matter.

\end{document}